\documentclass{article}
\usepackage{arxiv}
\usepackage{graphicx}
\usepackage{natbib}
\usepackage{hyperref}
\usepackage{listings}
\usepackage{xcolor}

\definecolor{backcolour}{rgb}{0.95,0.95,0.92}

\lstdefinelanguage{json}{
    basicstyle=\ttfamily\footnotesize,
    commentstyle=\color{blue},
    stringstyle=\color{purple},
    numbers=left,
    numberstyle=\tiny\color{gray},
    stepnumber=1,
    numbersep=5pt,
    showstringspaces=false,
    breaklines=true,
    frame=single,
    backgroundcolor=\color{backcolour},
    tabsize=2,
    string=[s]{"}{"},
    comment=[l]{:\ "},
    morecomment=[l]{:"}
}

\lstdefinelanguage{sparql}{
    basicstyle=\ttfamily\footnotesize,
    commentstyle=\color{olive},
    stringstyle=\color{blue},
    keywordstyle=\color{purple},
    numbers=left,
    numberstyle=\tiny\color{gray},
    stepnumber=1,
    numbersep=5pt,
    showstringspaces=false,
    breaklines=true,
    frame=single,
    backgroundcolor=\color{backcolour},
    tabsize=2,
    string=[s]{"}{"},
    morekeywords={*,PREFIX,SELECT,DISTINCT,WHERE,FILTER,BIND,AS,UNION},
    comment=[l]{\#\ },
    string=[s]{"}{"},
    morestring=[s]{<http}{>},
    morestring=[s]{<ipfs}{>},
    morestring=[s]{<did}{>}
}

\begin{document}

\title{Semantics and Non-Fungible Tokens for Copyright Management on the Metaverse and Beyond}


\author{ \href{https://orcid.org/0000-0003-2207-9605}{\includegraphics[scale=0.06]{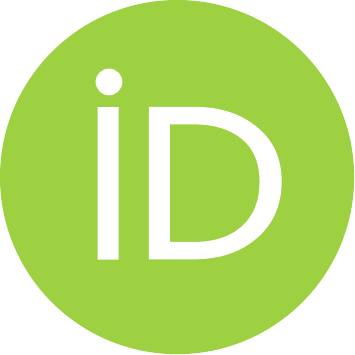}\hspace{1mm}Roberto García}\thanks{Corresponding author} \\
	Computer Science and Industrial Engineering\\
	Universitat de Lleida\\
	Víctor Siurana, 1, 25003 Lleida, Spain \\
	\texttt{roberto.garcia@udl.cat} \\
	\And
	\href{https://orcid.org/0000-0002-8495-2505}{\includegraphics[scale=0.06]{orcid.pdf}\hspace{1mm}Ana Cediel} \\
	Department of Public Law\\
	Universitat de Lleida\\
	Víctor Siurana, 1, 25003 Lleida, Spain \\
	\texttt{anna.cs@udl.cat} \\
	\And
	\href{https://orcid.org/0000-0001-9124-1464}{\includegraphics[scale=0.06]{orcid.pdf}\hspace{1mm}Mercè Teixidó} \\
	Computer Science and Industrial Engineering\\
	Universitat de Lleida\\
	Víctor Siurana, 1, 25003 Lleida, Spain \\
	\texttt{merce.teixido@udl.cat} \\
	\And
	\href{https://orcid.org/0000-0001-6304-9635}{\includegraphics[scale=0.06]{orcid.pdf}\hspace{1mm}Rosa Gil} \\
	Computer Science and Industrial Engineering\\
	Universitat de Lleida\\
	Víctor Siurana, 1, 25003 Lleida, Spain \\
	\texttt{rosamaria.gil@udl.cat} \\
}


\renewcommand{\shorttitle}{Semantics and Non-Fungible Tokens for Copyright Management}

\hypersetup{
pdftitle={Semantics and Non-Fungible Tokens for Copyright Management on the Metaverse and Beyond},
pdfsubject={q-cs.NC, q-bio.QM},
pdfauthor={Roberto García, Ana Cediel, Mercè Teixidó, Rosa Gil},
pdfkeywords={metaverse, non-fungible token, copyright, social media, blockchain, ontology},
}

\maketitle

\begin{abstract}
 Recent initiatives related to the Metaverse focus on better visualisation, like augmented or virtual reality, but also persistent digital objects. To guarantee real ownership of these digital objects, open systems based on public blockchains and Non-Fungible Tokens (NFTs) are emerging together with a nascent decentralized and open creator economy. To manage this emerging economy in a more organised way, and fight the so common NFT plagiarism, we propose CopyrightLY, a decentralized application for authorship and copyright management. It provides means to claim content authorship, including supporting evidence. Content and metadata are stored in decentralized storage and registered on the blockchain. A token is used to curate these claims, and potential complaints, by staking it on them. Staking is incentivized by the fact that the token is minted using a bonding curve. The tokenomics include the resolution of complaints and enabling the monetization of curated claims. Monetization is achieved through licensing NFTs with metadata enhanced by semantic technologies. Semantic data makes explicit the reuse conditions transferred with the token while keeping the connection to the underlying copyright claims to improve the trustability of the NFTs. Moreover, the semantic metadata is flexible enough to enable licensing not just in the real world. Licenses can refer to reuses in specific locations in a metaverse, thus facilitating the emergence of creative economies in them.
\end{abstract}

\keywords{metaverse, non-fungible token, copyright, social media, blockchain, ontology}

\maketitle

\section{Introduction}

Creative industries have been dealing with profound changes since the introduction of digital technologies and communication networks, which facilitate content creation but also copying and uncontrolled distribution. Blockchain technologies reintroduce scarcity, so it is possible to manage unique assets and their ownership through Non-Fungible Tokens (NFTs) \cite{clark_nfts_2021}, but they do not guarantee authenticity.

Thus, serious problems with fake NFTs and plagiarism have emerged, where a new NFT collection of unique tokens points to copied pieces of content. Just like a copied painting with a forged certificate of authenticity, but much easier to fake. And the issue is even more acute in completely digital spaces like metaverses, where emerging creative economies might be hampered by the lack of some sort of authorship recognition.

Classical copyright law can help here, but it should be integrated and scaled to these emerging digital spaces while benefitting from the new opportunities offered by these technologies. Our proposal, presented in this paper, is CopyrightLY, a decentralized application\cite{Cai_Wang_Ernst_Hong_Feng_Leung_2018} that leverages blockchain \cite{walport_distributed_2016} and semantic web \cite{shadbolt_semantic_2006} technologies to facilitate copyright management. A Proof of Concept (PoC) has been developed as part of the European Next Generation Internet ONTOCHAIN project \footnote{\url{https://ontochain.ngi.eu}}, whose aim is to create an ecosystem of blockchain-based knowledge management solutions.

In this context, CopyrightLY contributes a content ownership and a copyright management layer. One of its more direct usage scenarios, detailed in Section~\ref{sec:scenario} is being integrated with existing social media platforms and allowing content creators to explore ways to exploit their novel media beyond those made possible by those platforms, especially through NFTs and even across metaverses.

CopyrightLY's PoC provides a working blockchain-based copyright registry, where creators can make authorship claims linked to a hash of their content and a timestamp. They can then strengthen their claims by adding supporting evidence, like pictures of the creative process, witnesses recordings or proofs of previous appearances of the creation on social media platforms. CopyrightLY also provides the means to curate authorship in a scalable way, through token-based incentives and the wisdom of the crowds. As a last resort, all accumulated evidence, both for authorship claims and complaints, can be used in the context of arbitration systems or even in court.

Additionally, CopyrightLY showcases the use of Semantic NFTs. And NFT is a unique and non-interchangeable token linked to metadata whose ownership is recorded in a blockchain. Blockchain transactions are used to transfer ownership and metadata usually points to the media the NFT represents. Semantic NFTs have their metadata enriched using semantics and formal knowledge representation languages based on ontologies. CopyrightLY uses them as the mechanism to license the reuse of the registered content. This way, a Semantic NFT is linked to an authorship claim and an unambiguous representation of the terms licensed to the NFT owner. Moreover, the semantic metadata can be used to automate copyright reasoning and assist users during content reuse. The ontologies being used are flexible enough to accommodate reasoning in different domains, including licensing in a metaverse.

The rest of the paper is organized as follows. Section~\ref{sec:soa} presents the state of the art regarding blockchain technologies for copyright management, including how it might be integrated into a metaverse architecture. Then, Section~\ref{sec:approach} presents the proposed approach focusing on its architecture. Highlights of CopyrightLY’s implementation are provided in Section~\ref{sec:implementation}, including token-based incentives to curate authorship claims and the use of Semantic NFTs to unambiguously capture the licensing terms. Finally, Section~\ref{sec:conclusions} presents the conclusions and Section~\ref{sec:futurework} the future work.

\subsection{Sample Usage Scenario} \label{sec:scenario}

\begin{figure}[!htb]
 \centering
 \includegraphics[width=\linewidth]{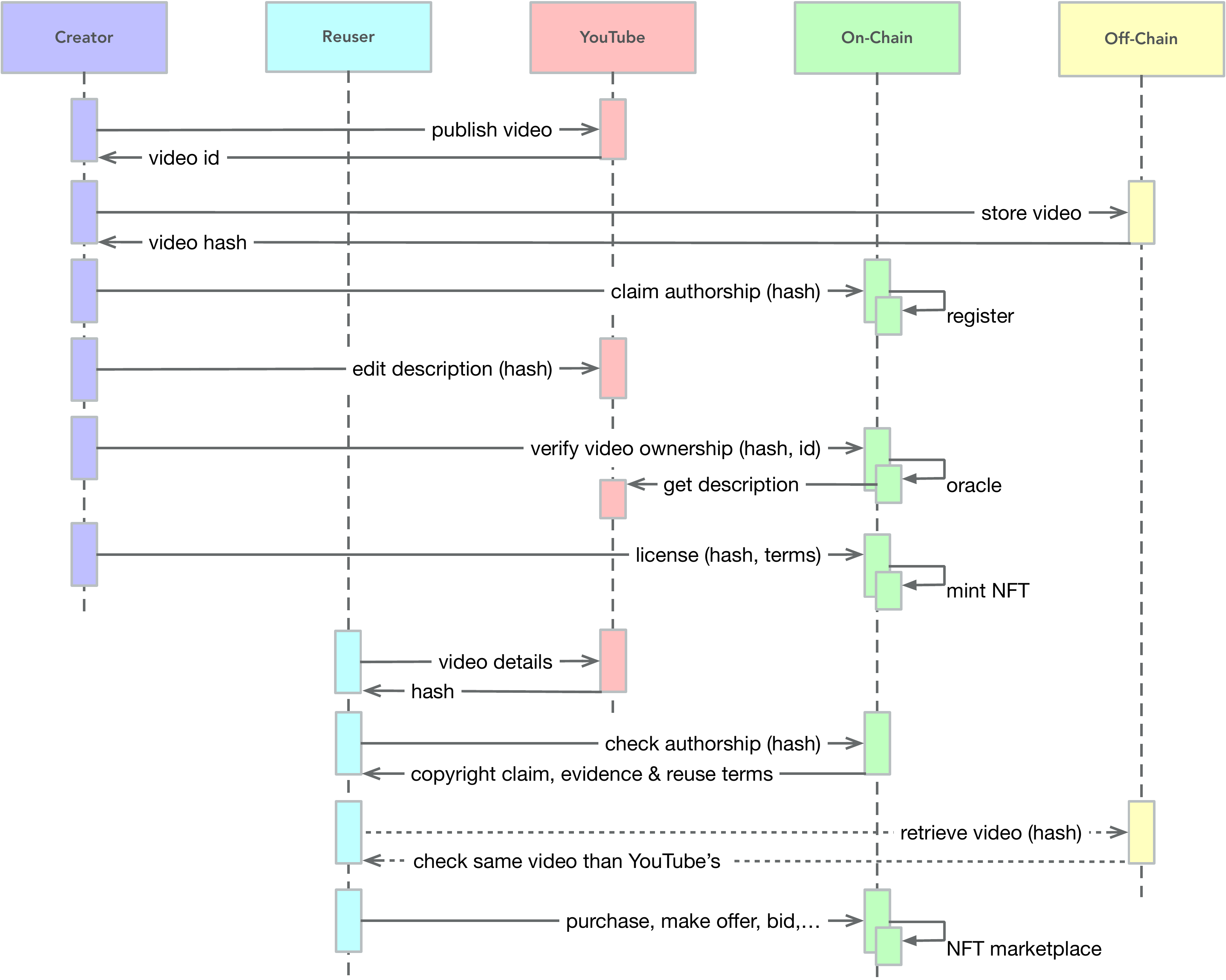}
 \caption{Typical usage scenario for a YouTube video creator using CopyrightLY as an alternative monetization mechanism through Non-Fungible Tokens.}
 \label{fig:scenario}
\end{figure}

A typical CopyrightLY usage scenario is depicted in Figure~\ref{fig:scenario} using a use case diagram. In this case, a YouTube video creator is willing to explore alternative ways to monetize her content. In addition to publishing the video on YouTube, she registers her authorship claim by uploading the video file to an immutable data store (off-chain) and registers its content hash using a blockchain transaction (on-chain). This produces a time-stamped and immutable authorship claim. The claim links the content through its hash (which can be used to retrieve the content), the blockchain identity used by the creator (linked to a private key used to sign the claim) and a timestamp.

The creator can then strengthen her claim by linking it to her YouTube video, and the social media profile and reputation that she might have built there. As shown in Figure~\ref{fig:scenario}, ownership of the social media platform version is proofed by including the authorship claim content hash into the YouTube video description. The presence is then checked through a blockchain oracle \cite{kochovski_oracles_2019}, a trusted service that brings data from other information systems to the blockchain. In this case, the oracle consumes the YouTube API, verifies that it includes the authorship claim hash and registers on-chain the YouTube video as supporting evidence for the claim. The logic is similar to the one used to verify that someone has a particular e-mail, sending to it a link that, when followed, is used to prove control over that e-mail account.

At this point, the content creator can start monetizing her claimed copyright through Non-Fungible Tokens, which are linked to the authorship claim, YouTube evidence and licensing terms. By following the ERC-721 standard \footnote{\url{https://eips.ethereum.org/EIPS/eip-721}}, NFTs can be integrated into existing NFT marketplaces and become tradable right away, so they can be offered for a fixed price or through different kinds of auctions. They can be also reused in the context of blockchain-powered metaverses, for instance, displayed on a virtual gallery. And potential purchasers or users can trace the NFT or YouTube video back to the underlying copyright claim and reuse terms to verify them.

\section{State of the Art} \label{sec:soa}

Blockchain technologies are being applied to almost any conceivable domain, from logistics \cite{dobrovnik_blockchain_2018} to renewable energy production management \cite{mengelkamp_blockchain-based_2018}. In all cases, in addition to the technologies specific to the distributed ledger itself, developers face issues related to information management and integration similar to those faced by Web applications. In this regard, Semantic Web technologies are well-positioned to solve this kind of issue, and have been even used to model the key concepts in one of the main distributed ledgers, Ethereum’s EthOn \cite{pfeffer_ethon-ethereum_2018}.

However, attempts to apply blockchain technologies for completely decentralized copyright registration and monetization are quite recent and there is little literature regarding proposals similar to CopyrightLY’s. One similar proposal \cite{kripa_2021}, which also aims to facilitate copyright protection of social media content, proposes a blockchain framework with smart contracts to protect social media contents using IPFS (InterPlanetary File System), a decentralized and content-addressable peer-to-peer hypermedia storage protocol \cite{benet_ipfs_2014}. Content uploaded to IPFS is securely stored using a secret sharing scheme. This is combined with a robust hash for images, a method of hashing images that is resistant to modification, rotation, and colour alteration. The objective is to make it possible to detect near copies and block the registration of images that might be copies of previously registered ones. Unfortunately, no further information is provided about the robustness of this algorithm and its implications from a legal standpoint when infringements are not properly detected or uninfringing content is considered otherwise. Moreover, though the use of smart contracts is mentioned as the way to ask permission to reuse registered images, no further details are provided about these smart contracts or the way reuse terms are negotiated and then stored on-chain to provide trust in those agreements.

Another similar proposal \cite{dobre_blockchain-based_2020} is a system to fight intended or accidental image copyright infringement on social media platforms, mainly when professional images are used to increase the impact of posts. Photographers can use it to register their photos and re-users can use it to check if the image they want to use is copyright protected or not. The main result of \cite{dobre_blockchain-based_2020} is an algorithm that can extract a signature that is resistant to different levels of JPEG compression. The signature is stored on the blockchain along with the identification data of the copyright owner. It can be then used to detect copies when someone tries to register the same or a similar image, as determined by the algorithm. The algorithm can be also used by re-users to check if an image is already registered. In that case, the system allows the purchase of the right to use the photo. However, \cite{dobre_blockchain-based_2020} does not provide mechanisms to deal with situations when the registration that is considered a copy is in fact the original one, giving rise to potential complaints. Moreover, the paper focuses just on the detection mechanism and few details are provided about how registrations are stored on-chain, or how reuses are negotiated and managed using the smart contracts that are mentioned. Approaches based on watermarking algorithms that link the copyright information with the content file \cite{Natgunanathan_Praitheeshan_Gao_Xiang_Pan_2022} experience the same kind of limitations, as there is no prior mechanism to ensure that the watermark is embedded by the actual creator or rights holder.

As will be detailed in the next section, the main innovation of CopyrightLY, and what distinguishes it most from blockchain-based copyright management proposals like the previous ones, is that it does not try to enforce what is an infringing authorship claim. As shown by previous experiences with Digital Rights Management (DRM) systems \cite{lessig_code_1999}, a rigid and technologically enforced approach might put this kind of systems outside copyright law. For instance, blocking fair use or categorising as copyright infringement registration attempts that might be done by the original creator.

Finally, in the yet scarce literature about the metaverse and blockchain technologies, we have been unable to find any previous work directly connected with copyright management. However, a proposal about metaverse architecture \cite{Duan_Li_Fan_Lin_Wu_Cai_2021} is a valuable reference that helps us situate CopyrightLY's contribution in the context of their proposed three main layers of a metaverse:
\begin{itemize}
 \item \textbf{Infrastructure}: this is the base layer on top of which the metaverse is deployed. It is based on a computation and communication infrastructure, including blockchain technologies, in our case Ethereum, and storage, for CopyrightLY is decentralised storage provided by the InterPlanetary File System (IPFS).
 \item \textbf{Interaction}: this layer deals with immersive user experiences, and interfaces to shape the metaverse and digital twins. CopyrightLY does not make any contribution in this regard and, when integrated into an existing metaverse, it just makes use of its interactive features. For instance, in the implementation example provided in Section~\ref{sec:semanticnft}, Copyright is integrated into the Voxels metaverse\footnote{\url{https://www.voxels.com/}}.
 \item \textbf{Ecosystem}: this is the metaverse layer where CopyrightLY focuses its contributions. The following aspects are highlighted for the three main components of the Ecosystem layer:
       \begin{itemize}
        \item \textbf{User Generated Content (UGC)}: CopyrightLY helps deal with UGC by managing its underlying copyright. UGC can be of any kind, from metaverse specific to social media like YouTube videos. In all cases, after its registration through curated authorship claims, they can be licensed and integrated as NFTs into marketplaces and metaverses for their reuse.
        \item \textbf{Economics}: CopyrightLY's NFT provide the ownership and trading mechanisms that enable open creator economies in the metaverse and beyond. Regarding the governance of the economics layer, CopyrightLY also features a dedicated token with economic value that drives the incentives for authorship claims curation, as detailed in Section~\ref{sec:incentives}.
        \item \textbf{Artificial Intelligence}: CopyrightLY provides mechanisms to enrich NFTs with machine actionable copyright information. They are used to unambiguously state the reuse terms granted through token ownership, which can be fed into reasoners to assist the user during content reuse. For instance, as detailed in Section~\ref{sec:semanticnft}, the reasoner can determine if content can be displayed in a specific metaverse location based on the terms accompanying the corresponding NFT.
       \end{itemize}
\end{itemize}

\section{Approach}\label{sec:approach}

The most distinctive characteristic of the proposed approach is to not rely beforehand on automated mechanisms to determine infringement, like the algorithms just mentioned in the State of the Art for robust hashing or fingerprinting. The content hashes used by CopyrightLY are just to permanently link the authorship claim registered on-chain with the content of the claimed creation stored on decentralised and immutable storage. This way it is possible to track the content linked to a claim and even retrieve it. Consequently, CopyrightLY just detects exact copies, because they generate the same content hash when making an authorship claim. Even in that case, as detailed in the following sections, the potentially original creator can proceed though in this case to file a complaint. Copyright is a social construct and, thus, it is difficult or maybe impossible to fully automate it and create an algorithm capable of discerning between originals and copies.

This approach was already introduced in \cite{garcia_copyrightly_2021}, where we depicted just its preliminary architecture that is fully developed in this paper in Section~\ref{sec:architecture}. We complement now the proposed approach with an incentives system to curate the pool of authorship claims and complaints made through CopyrightLY as detailed in Section~\ref{sec:incentives}. Additionally, it is important to note that all the accumulated claims, complaints and evidence are registered on-chain and can be used in case of litigation and admissible as evidence in a judicial proceeding, together with expert assessment, like in the TikTok vs Baidu case \cite{michalko_how_2020}.

Consequently, the proposal is aligned with copyright law and courts, relying on them as last resort. However, to improve its scalability, it also integrates incentives and arbitration mechanisms following a blockchain-based decentralized approach. Grounding the system on copyright also helps deal with one of the main issues in NFTs state of the art, issues derived from their disconnection from the underlying copyrights \cite{schmalfeld_nft_2021}. In addition to being grounded on copyright ownership, NFTs should also clearly state the licensing terms and conditions to become more than mere transfers of ownership on just a piece of metadata \cite{mezei_rise_2021}, as detailed in Section~\ref{sec:semanticnft}.

CopyrightLY is based on the combination of a set of Ethereum \cite{antonopoulos_mastering_2018} blockchain smart contracts that manage authorship claims, complaints, evidence and licensing terms. These smart contracts are connected with decentralized storage, where content and metadata are stored to avoid the prohibitive costs of storing them on-chain. Moreover, there are a set of blockchain oracles \cite{kochovski_oracles_2019} that connect on-chain smart contracts to off-chain APIs of the main social media platforms, like Twitter, Facebook or YouTube.

There is a web front-end that facilitates users' interaction with on-chain logic and state. All blockchain read operations are carried out through an API that provides direct access to a representation of the on-chain state. This way, users can interact as they will do with any other web applications not based on blockchain technologies. For interactions modifying blockchain state, end-users use an Ethereum wallet, as a browser extension or on their mobile devices.

The wallet holds the public-private key pair used to sign on-chain transactions. The public key is pseudo-anonymous but, following a Self-Sovereign Identity approach \cite{preukschat_ssi_2021}, users can hold credentials in their wallets that can be presented to requesting parties when stronger identification is required. These credentials range from social media profiles, which can be openly disclosed and shown in the creator's profile, to legal identity credentials that are kept private and selectively presented. For instance, in case of litigation to better support on-chain evidence in court.

As previously introduced, NFTs are used as the monetization mechanism through with reuse terms are licensed and traded. In addition to helping keep track of ownership, NFTs are enhanced with semantic metadata to capture the terms being transferred and the relationship between the token and the copyright claim. Semantic metadata is based on existing vocabularies like Schema.org \cite{guha_schema.org:_2016} and the Copyright Ontology \cite{garcia_infsys_2010}. A rights module enhanced with semantic technologies and reasoning capabilities fetches data from NFTs and combines it with ownership information registered on-chain. It uses it to respond to rights-related issues, like checking if an intended reuse is authorized by any owned licenses or finding available licensing NFTs that might provide it. The following subsection details the architecture developed to implement this approach.

\subsection{Architecture Details} \label{sec:architecture}

\begin{figure}[!htb]
 \centering
 \includegraphics[width=\linewidth]{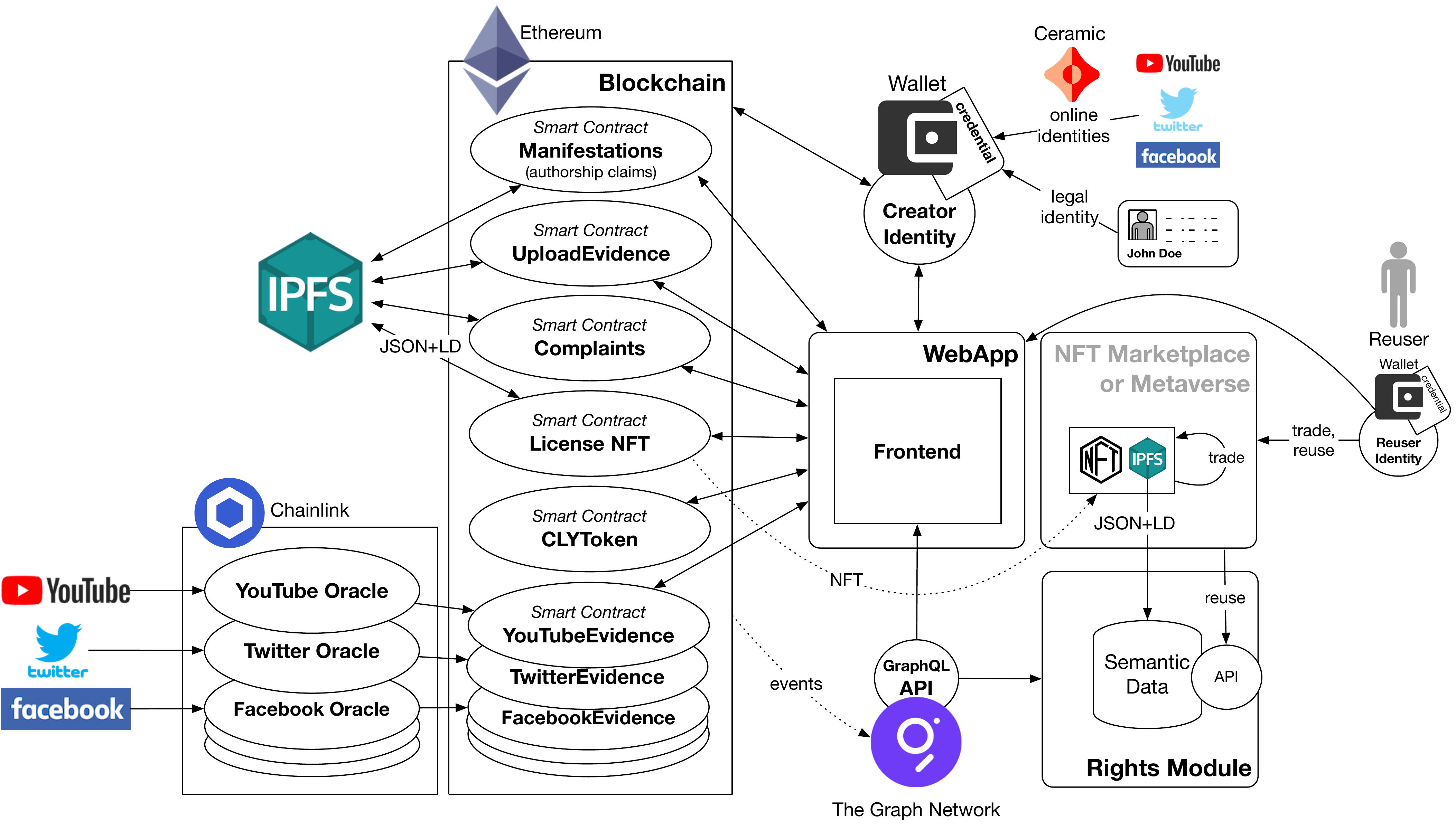}
 \caption{Proposed architecture for CopyrightLY, including a layer of smart contracts operating on-chain (centre), a set of oracles bringing off-chain data on-chain (left) and a front-end and an API interfaces also managing user credentials (right).}
 \label{fig:architecture}
\end{figure}

The central part of CopyrightLY’s architecture is on-chain and based on a set of smart contracts, as shown at the centre of Figure~\ref{fig:architecture}. The smart contracts take care of registering Manifestations (i.e. authorship claims) and Complaints (to denounce authorship claims as potentially fraudulent). Authorship claims and complaints are supported with evidence. The simplest kind of evidence is based on files uploaded to decentralized storage, managed by the UploadEvidence smart contract that registers and links them to manifestations and the accounts triggering the transaction. All the previous smart contracts use decentralized storage based on IPFS to store manifestations, complaints or evidence content (data, documents or any kind of media can be uploaded).

Another way of supporting claims with evidence is through the oracles responsible for asserting on-chain the off-chain information about ownership of social media content (YouTube, Twitter or Facebook). For each social media platform, the oracle verifies information about social media assets through the corresponding APIs, like YouTube’s API to verify that a predefined identifier has been added to a video description. This mechanism allows checking that an on-chain account has control over the corresponding off-chain social media assets. The oracles infrastructure is based on ChainLink \footnote{\url{https://chain.link}}, a decentralized network giving applications access to trusted off-chain computation and data.

Finally, there are a couple of smart contracts dealing with tokens. The License NFT contract implements an ERC-721 Non-Fungible Token, i.e. a kind of token that is unique and not interchangeable. It can be minted by a creator to monetize her copyright. The NFT metadata, in addition to the standard attributes like name or image, represents the reuse terms to be transferred to the future owner of the NFT, as further detailed in Section~\ref{sec:semanticnft}.

On the other hand, the CopyrightLY token (CLY) implemented by the CLYToken smart contract is an ERC-20 fungible token, interchangeable like money. This smart contract makes this token particular in the sense that it can just be minted, in exchange for Ethereum's native cryptocurrency Ether, to be staked on an existing authorship claim or complaint. If unstaked, it is burned back into Ether. This token implements the incentive mechanism to curate authorship claims and complaints, as detailed in Section~\ref{sec:incentives}.

Three off-chain components are interacting with all this on-chain infrastructure. First of all, a Web front-end through which creators and reusers interact with on-chain logic and state. They do so using Ethereum wallets that link the accounts used to interact with the blockchain with verifiable credentials following a decentralized Self-Sovereign Identity approach \cite{preukschat_ssi_2021}. This component stores the credentials issued to users' on-chain identities in a private and self-sovereign way. Users can request that third parties issue these credentials after a verification process. In the case of CopyrightLY, the objective is to have both credentials about control of social media profiles and legal identities (required in case of litigation).

Additionally, there is a back-end that monitors blockchain events, keeps track of the on-chain state and provides a better user experience because it does not require that users connect directly to the blockchain. This component is based on The Graph\footnote{\url{https://thegraph.com}}, an indexing protocol for querying networks like Ethereum. It makes a model of the on-chain state for a set of smart contracts by monitoring the events they generate, which is then available through a GraphQL \cite{hartig_graphql_2018} API.

The last off-chain component is the Rights Module, responsible for NFTs' semantic data exploitation and providing reasoning mechanisms based, fundamentally, on the Copyright Ontology. License NFTs can be automatically traded and integrated into existing NFT marketplaces and metaverses, as shown in Figure~\ref{fig:nftmarketplace}, as they are based on the ERC-721 standard. The Rights Module monitors minting and ownership change events through the GraphQL API. Moreover, it retrieves NFTs' semantic metadata from IPFS by following the token URI also defined by the NFT standard. All this information is stored in a semantic repository providing the reasoning and semantic query capabilities, available through a SPARQL \cite{ducharme_sparql_2013} API. Through this API, reusers can, for instance, check if an intended reuse is allowed by the NFTs they own in the context of a specific metaverse location. Further details about the implementation of this feature are provided in Section~\ref{sec:semanticnft}.

\section{Implementation Highlights} \label{sec:implementation}

The previous architecture has been implemented as a Proof of Concept in the context of the ONTOCHAIN NGI H2020 project. It is possible to follow the development process, get access to the source code and test the current version of the application from the project’s dashboard\footnote{\url{https://github.com/rhizomik/copyrightly}}. The dashboard also includes the automated tests used to evaluate CopyrightLY compliance with the requirements specification. The following subsections highlight some of the most relevant parts of the project, including the incentive mechanisms to curate authorship claims and the implementation of licensing NFTs based on semantic technologies and ontologies, which enable copyright reasoning on top of NFTs ownership and tradeability mechanisms.

\subsection{Incentives to Curate Authorship Claims} \label{sec:incentives}

One of the biggest issues detected in the state-of-the-art proposals analyzed in Section~\ref{sec:soa} is that registering content on-chain does not provide any guarantee of its validity. It might be just the action of someone with access to a digital copy of a creation, but not necessarily the original creator. To deal with these false authorship claims, CopyrightLY leverages two incentive mechanisms. The first one is based on the fact that all claims and evidence are recorded on-chain and can be used in case of litigation. This is possible even in court, like in the TikTok vs. Baidu case \cite{michalko_how_2020}, and disincentivizes false claims. However, it just works in the longer term and when some identifying information has been provided to strengthen the claim.

The second incentive mechanism aims to work on the shorter-term, scale thanks to the wisdom of the crowds and incentive stronger authorship claims and complaints, supported by evidence and some identifying information like social media accounts linked to the user identity. The approach is based on a Token Curated Registry \cite{hassanien_tcrs_2021}, where a token with economic value, the CLY token, is used as an incentive mechanism \cite{zhou_incentives_2021} to curate the most reliable manifestations for authorship claims and complaints in a decentralized way.

\begin{figure}[!htb]
 \centering
 \includegraphics[width=\linewidth]{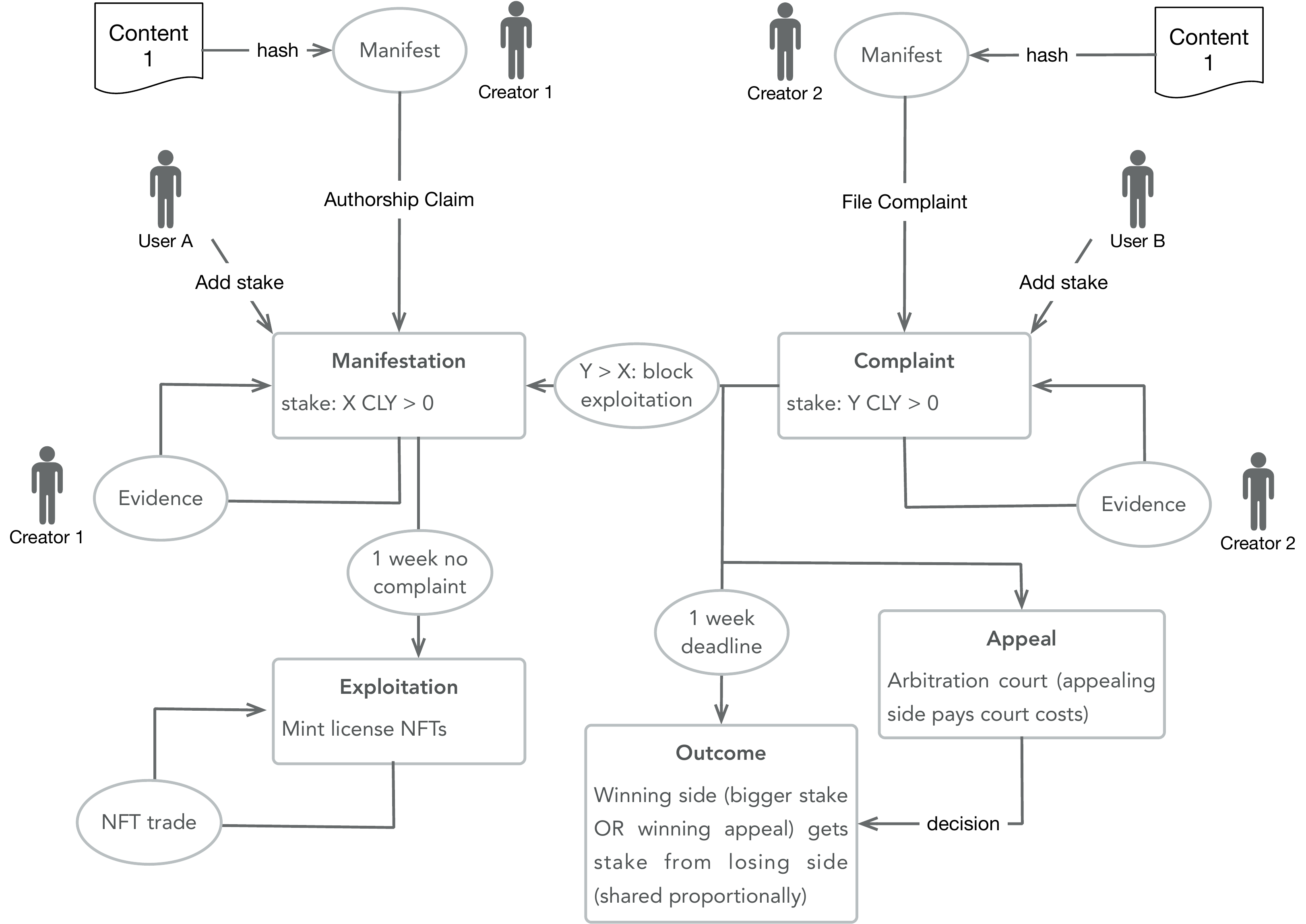}
 \caption{Tokenomics of the CLY token to incentive the curation of authorship claims.}
 \label{fig:tokenomics}
\end{figure}

The mechanism requires creators to stake an amount of the CLY token together with their authorship claims, as shown in the upper left part of Figure~\ref{fig:tokenomics}, which depicts CLY tokenomics. Following the "skin in the game" heuristic \cite{taleb_skin_2014}, creators risk losing their stake when other creators complain about existing claims, as shown in the upper right part of the same figure. Like original claimers, they can also provide supporting evidence to convince other users to support their position with additional stake. In case of conflicting claims, stakes cannot be removed till it is solved and the side with more stake wins the stake of the other side and shares it proportionally to their stake.

Additionally, filing a complaint on an existing authorship claim blocks the ability of its creator to continue monetizing it until the conflict is solved. Creators can start minting NFTs to license their copyright after staking some CLY and a time has passed without complaints, for instance, one week. On the other hand, to avoid spam complaints that look to block monetization opportunities, they become blocking just if the amount of CLY staken on the complaint exceeds that on the challenged manifestation.

To avoid players with big CLY token stakes dominating the game, an appeal mechanism based on an external arbitration court and decentralized justice \cite{aouidef_kleros_2021} is also considered. In this case, any of the sides can appeal before the deadline to solve a complaint. The appealing side brings the case to a decentralized arbitration system like Kleros\footnote{\url{https://kleros.io}}. A court is opened on-chain with links to all evidence accumulated by both sides plus any identifying credentials that the parties might provide. Systems like Klerors are open to jurors that are invited to decide on cases. There is also a token, in the case of Kleros it is called Pinakion (PNK), which jurors have to stake and which is used to incentive participation and avoid attacks \footnote{\url{https://medium.com/kleros/why-kleros-needs-a-native-token-5c6c6e39cdfe}}.

Finally, to motivate participation in this staking game, the CLY token has an inflationary behaviour \cite{wang_tcr-inflation_2019} tied to its use. The token price, in Ether, follows a bonding curve \cite{zargham_bondingcurve_2020} and grows with the token supply. Consequently, if the CLY token succeeds and is used more, which requires minting it to increase the token supply, its price also grows. Early adopters can thus benefit from future players entering the game and unstake and burn their CLY for an amount of Ether bigger than what they originally invested. However, to avoid mere speculation, it is important to note that CLY can be staked but not held. And stakers should take care of doing so on trustful claims that do not put them at risk.

\subsection{Semantic NFTs and Copyright Reasoning} \label{sec:semanticnft}

The main contribution of CopyrightLY is that authorship claims can be monetized using Semantic NFTs, which solve the main issues identified in the state-of-the-art Section~\ref{sec:soa}. First of all, NFTs are linked to manifestations and thus rooted in an authorship claim accompanied by evidence, a commitment in the form of staked CLY and potentially curated as previously detailed.

Additionally, NFTs' metadata define the scope of the licensed rights, which are modelled using standard semantic technologies, like the Web Ontology Language (OWL) \cite{OWL2Primer_2012}, and based on the Copyright Ontology \cite{garcia_infsys_2010}. They provide the necessary and unambiguous building blocks that also make the terms machine-actionable.

As shown in the architecture, Figure~\ref{fig:architecture}, the License NFT smart contract registers them on-chain and stores their metadata on decentralized storage based on IPFS. The licensing terms are serialized as semantic data based on the JSON-LD\footnote{\url{https://json-ld.org}} standard and stored on IPFS resulting in a content hash. This hash is used as the NFT token URI, used to point to the token metadata. The token URI is immutable so it can be verified that the terms have not changed after the token is minted.

The JSON-LD serialisation is configured so the ERC-721 standard metadata fields (\emph{name}, \emph{description}, \emph{external\_link}, \emph{image} and \emph{animation\_url}) are available as plain JSON. This way, Semantic NFTs' metadata is compatible with existing NFT marketplaces. For instance, Listing~\ref{lst:jsonld} shows the metadata for an NFT with the name "Reuse license for 'Copyright Blockchain'", which corresponds to a piece of video art generated by an Artificial Intelligence, concretely VQGAN+CLIP \cite{Cetinic_She_2022}. Then, Figure~\ref{fig:nftmarketplace} shows that Semantic NFT in an existing NFT marketplace, so ERC-721 interoperability with existing metaverses and marketplaces is retained by Semantic NFTs.

The metadata includes a \emph{description} field that describes the licensing terms as plain text for user consumption. The rest of the metadata is tied to different vocabularies and ontologies to enable an unambiguous and rich representation of the terms, which makes them machine-actionable. First of all, the Schema.org vocabulary \cite{guha_schema.org:_2016} is used for the NFT standard fields that have the same and meaning in Schema.org, i.e. \emph{name}, \emph{description} and \emph{image}. The rest of the metadata is responsible for capturing the license terms that grant certain uses of the copyrighted work to the NFT owner.

For the copyright-related part, terms from the Copyright Ontology\footnote{\url{https://rhizomik.net/ontologies/copyrightonto}} are used. They are tied to the \emph{cro} prefix in the \emph{@context} at the beginning of the JSON-LD serialization, Listing~\ref{lst:jsonld} line 4. Overall, the Copyright Ontology defines the following models:

\begin{itemize}
  \item {\textbf{Creation Model}}: the different shapes of copyright creations along their lifecycle, i.e. \textit{Manifestation}, \textit{Performance}, \textit{Recording}.... In the case of Semantic NFTs, authorship claims are \textit{Manifestations}, i.e. expression of the work into some kind of object (digital or not) that can be used to claim copyright.
  \begin{itemize}
    \item {\textbf{Example}}: for the sample Semantic NFT in Listing~\ref{lst:jsonld}, this corresponds to lines 30-33, where it is defined the manifestation for which actions are granted by the NFT license.
  \end{itemize}
  \item {\textbf{Actions Model}}: captures the copyright actions moving creations along their lifecycle (\textit{Manifest}, \textit{Perform}, \textit{Communicate}...) together with actions' dimensions (\textit{who}, \textit{what}, \textit{when}, \textit{where}...) and other management actions like \textit{Agree} or \textit{Disagree}. The approach is similar to Schema.org Actions \cite{brickley_announcing_2014} and many of the Copyright Ontology terms are aligned or directly reuse Schema.org ones.
  \begin{itemize}
    \item {\textbf{Example}}: the actions modelled in the Semantic NFT example are \textit{Agree} and \textit{MakeAvailable}, lines 11 and 23 respectively. Both are connected with \textit{who} performs the action. In the first case, the creator granting the license represented as a Decentralized Identifier (DID\footnote{\url{https://www.w3.org/TR/did-core/}}) for the Ethereum account used by the creator as shown in line 19. In the second case, anyone owning the license NFT can make the associated content available. The NFT is also identified using a DID, shown in lines 10 and 27. The NFT metadata also details other conditions for the reuse, including a start and end time (lines 24 and 25), an instrument (line 36, which points to the Voxels metaverse Web application) or the authorised territories (lines 39 and 40, Voxels' neighbourhood "The Center" and the "Vibes" island).
  \end{itemize}
  \item {\textbf{Rights Model}}: this part of the Copyright Ontology represents the legal constructs regulating what actions are favoured or restricted. Different legal systems can be represented, from generic ones geared towards worldwide harmonization like those proposed by the WIPO \cite{wipo_wipo_2004} to those in particular legal regimes. The Semantic NFT does not use any term from the Rights Model because what is licensed are specific actions and no rights are transferred.
\end{itemize}

\begin{figure}[!htb]
 \centering
 \includegraphics[width=\linewidth]{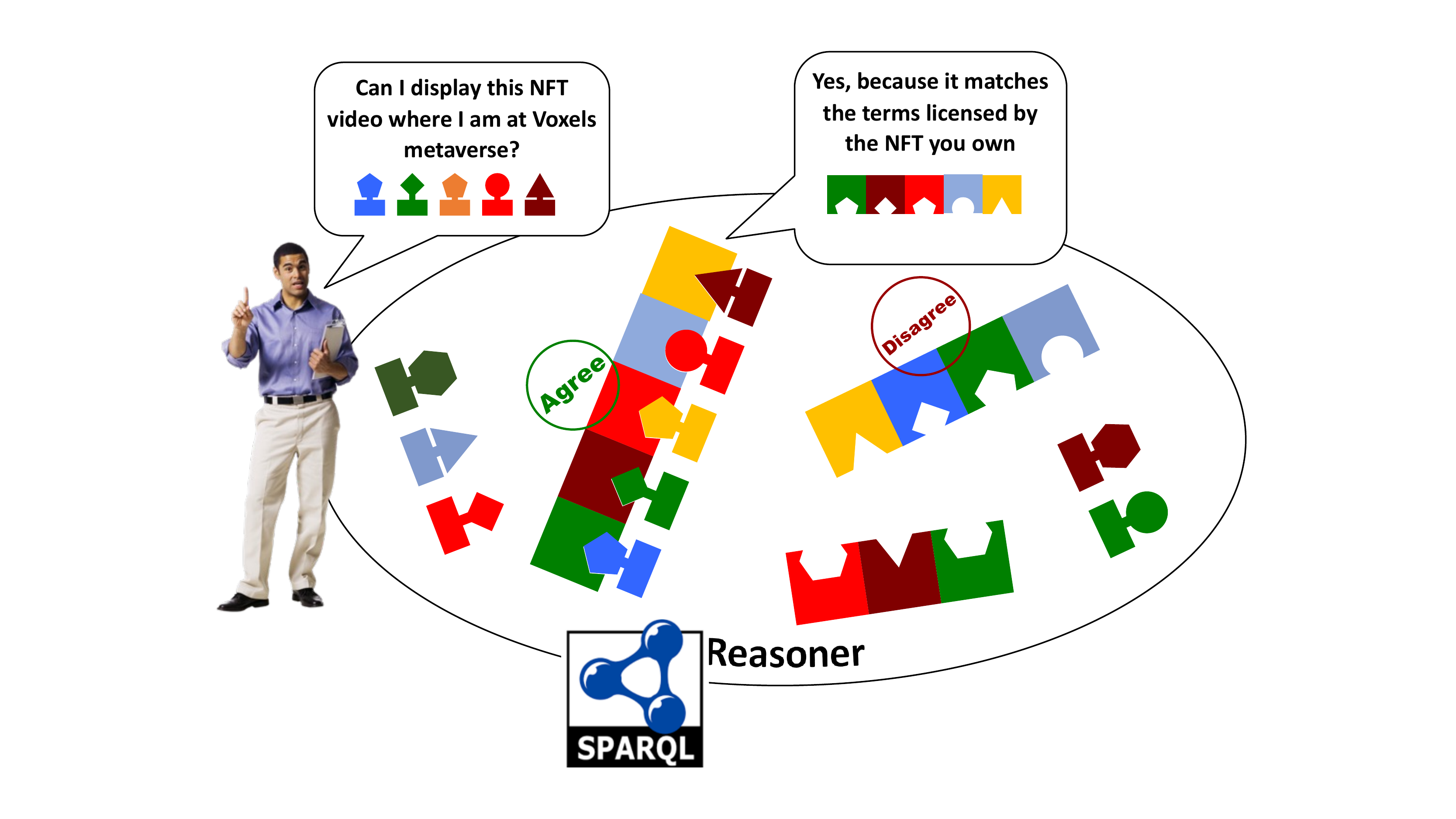}
 \caption{Illustration of how copyright reasoning can assist re-users to determine if and intended reuse is allowed by the NFT licenses they own.}
 \label{fig:reasoning}
\end{figure}

Based on the semantic data in NFTs, and in the context of the Rights Module introduced as part of the architecture in Section~\ref{sec:architecture}, it is possible to make the licensing terms machine-actionable. The rights module fetches the semantic data from minted NFTs and monitors on-chain events to track ownership. All this information is combined into a semantic repository with reasoning and semantic query capabilities. Reasoning has been implemented using SPARQL \cite{ducharme_sparql_2013} queries that check if the different dimensions of an intended reuse (who, what, when, where...) do fit within the licensing terms as illustrated in Figure~\ref{fig:reasoning}.

Matches can be direct, like that the content or data to be reused has a particular identifier (or content hash), but they can also be indirect and require some reasoning. For instance, if the territory where the intended action would take place is part or not of those where it is authorized. The SPARQL queries implementing copyright reasoning are hidden inside the Rights Module, which offers an API providing higher-level operations, like checking if an intended reuse action is allowed by the terms licensed to the user making the request.

\textbf{Example}: An example of a SPARQL query used by the Rights Module is presented in Listing~\ref{lst:sparql}. This query is used to check if a given user is authorised to reuse some specific content in a specific metaverse location, in this case, some specific coordinates of the Voxels metaverse. Based on the metadata for the NFT in Listing~\ref{lst:jsonld}, and if the intended reuser is the owner of the NFT, it will respond affirmatively to the reuse request and include a link to the agreement granting it, i.e. the DID of the license NFT. In this particular case, the Voxels metaverse location where the reuse is intended is coordinates 6510W,100N\footnote{\url{https://www.voxels.com/play?coords=W@6510W,100N}}, which correspond to Voxels' parcel 4962\footnote{\url{https://www.voxels.com/parcels/4962}} named "Generative Artworks Gallery". Though the NFT metadata explicitly authorises the reuse for Voxel's The Center\footnote{\url{https://www.voxels.com/neighborhoods/the-center}} neighbourhood and the Vibes island\footnote{\url{https://www.voxels.com/islands/vibes}}, the query implements all the reasoning about how territories are organised in Voxels metaverse. All that information has been captured using Voxels API and modelled as an ontology available online\footnote{\url{https://rhizomer.rhizomik.net/datasets/voxels}}. When the license NFT and ownership information is combined with the ontology about Voxels' territories and their "part of" relationships, the Rights Module is capable of determining that coordinates "6510W,100N", i.e. (-65.1 1), are located on parcel 4962 which belongs to the Vibes island, so that is also an authorised location and the user can display the NFT there.

\section{Conclusions} \label{sec:conclusions}

As shown in the previous sections, CopyrightLY contributes a decentralised authorship and rights management layer that has been evaluated in the context of the ONTOCHAIN NGI H2020 project ecosystem. This decentralized application leverages blockchain and semantic web technologies and implements services that allow claiming content authorship and providing supporting evidence.

To disincentivize false claims, a Token Curated Registry has been implemented based on the staking of the CLY token on authorship claims. Claims can be challenged through complaints, which also require staking CLY, and conflicts are resolved based on the amount of CLY that each party has been capable of attracting from anyone willing to participate in this crowdsources process. Participation is incentivized by binding CLY's price in Ether, the main cryptocurrency of the Ethereum blockchain, to a bonding curve.  Thus, early adopters can experience a capital gain when they burn CLY back into Ether.

An additional contribution is that CopyrightLY makes it also possible to monetize content using Semantic Non-Fungible Tokens (NFTs) grounded on copyright claims and stating the licensing terms using semantic metadata based on the Copyright Ontology. This solves two of the main issues that the NFT market is experiencing, spam NFTs and lack of clarity regarding what is owned when an NFT is purchased.

License NFTs metadata captures the reuse terms in a machine-actionable way, so it is possible to use automated reasoning to assist users with intended reuses. Moreover, despite NFT metadata being enriched with semantics, they are still compatible with existing marketplaces and metaverses. This makes it possible, as shown through an example, to use reasoning to determine if a license NFT allows displaying the associated content at a particular metaverse location.

\begin{figure}[!htb]
 \centering
 \includegraphics[width=\linewidth]{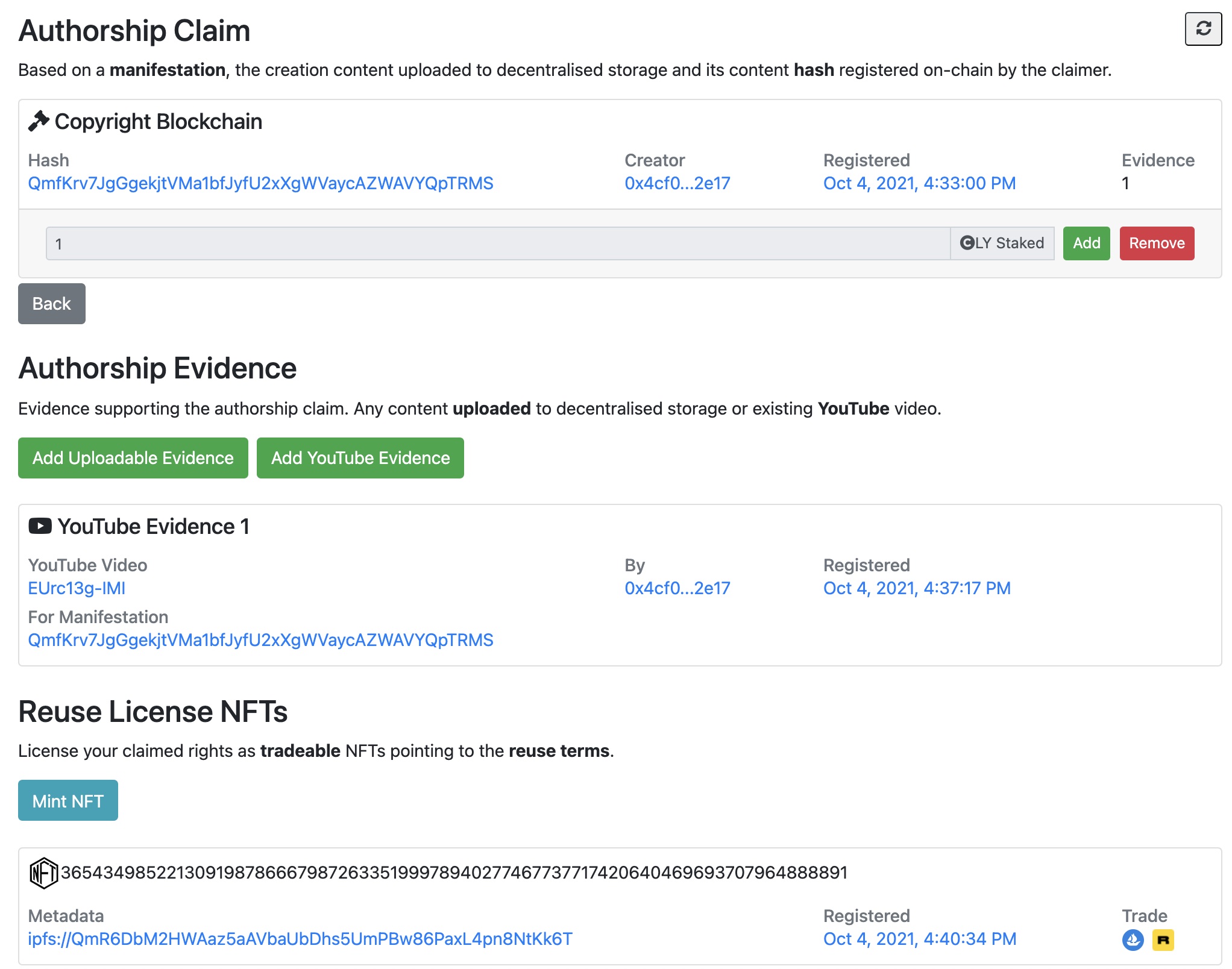}
 \caption{Authorship claim and supporting evidence linked to the Semantic NFT corresponding to the metadata in Listing~\ref{lst:jsonld}.}
 \label{fig:claim}
\end{figure}

\begin{figure}[!htb]
 \centering
 \includegraphics[width=\linewidth]{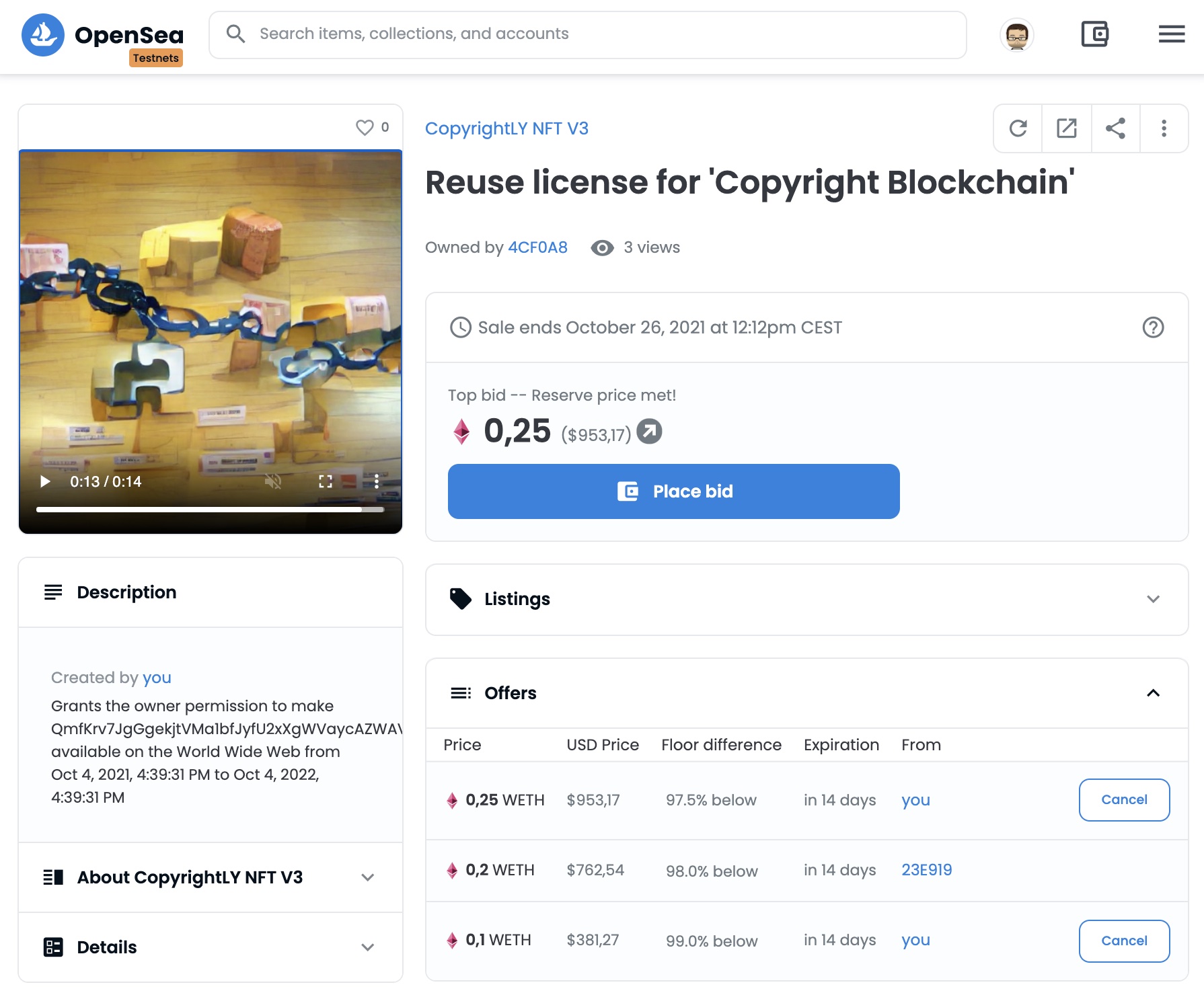}
 \caption{Semantic NFT traded on the OpenSea NFT marketplace, corresponding to the authorship claim shown in Figure~\ref{fig:claim}.}
 \label{fig:nftmarketplace}
\end{figure}

A prototype of the project is available online at\footnote{\url{https://copyrightly.rhizomik.net}} and deployed on the Rinkeby Ethereum testnet. Figure \ref{fig:claim} shows how an authorship claim looks in CopyrightLY, including a YouTube evidence and a licensing NFT. That NFT points to the terms and can be traded on NFT marketplaces, for instance OpenSea's testing marketplace\footnote{\url{https://testnets.opensea.io}}, as shown in Figure \ref{fig:nftmarketplace}.

\section{Future Work} \label{sec:futurework}

Though the tokenomics of the CLY token have been designed for both authorship claims and potential complaints, the current implementation only captures the authorship claims part. Future work is to complete the complaints’ part and implement the designed tokenomics mechanisms that settle the incentives to curate the set of authorship claims. This includes generating complaints when potentially infringing content is discovered and the incentives to do so.

To detect potentially infringing authorship claims, we anticipate using mechanisms beyond content hashes that detect just exact copies. However, as discussed in the state of the art Section~\ref{sec:soa}, mechanisms capable of detecting near-duplicates, like fingerprinting, should be used just as an aid for creators to find out false claims of their copyright. Then, they should go through the process of filing a complaint and providing evidence to attract the CLY stake to win the case. This might facilitate the curation process and, given the incentives, also attract other users to identify weak authorship claims and potential CLY stake rewards.

Additionally, though CopyrightLY licensing NFTs are tradeable on existing NFT marketplaces as it has been shown, existing marketplaces or metaverses miss the functionality to exploit the rich semantics captured by CopyrightLY NFTs metadata. Future work would be to develop NFT marketplaces and metaverse connectors capable of leveraging the copyright semantic metadata, for instance by reusing the Rights Module included in the project architecture.

Finally, work will continue by targeting potential users and other interested parties. For the moment, the following user profiles have been identified:
\begin{itemize}
 \item Social media creators that are willing to explore alternative monetization options, specially NFTs.
 \item Independent creators, especially those in developing countries lack access to the media industry or markets, even those unbanked.
 \item Media companies looking to license small creators to which current approaches do not scale, especially in developing countries.
 \item Marketplaces and metaverses looking for more trustful NFTs, linked to copyright claims and clearly stating the licensing terms.
\end{itemize}

CopyrightLY's user experience (UX) will be evaluated through the participation of potential users from previous profiles. Users will be requested to perform certain tasks under the supervision of UX experts that will evaluate the user experience. Moreover, post-test surveys will be used to evaluate user satisfaction and recommendations about future CopyrightLY improvements.

\bibliographystyle{ACM-Reference-Format}
\bibliography{biblio}

\appendix

\section{Semantic NFT Example}

\begin{lstlisting}[language=json, caption={JSON-LD serialisation of the metadata of a Semantic NFT granting its owner permission to make available the content registered by an authorship claim}, label=lst:jsonld]
{
  "@context": {
    "@vocab": "https://schema.org/",
    "cro": "https://rhizomik.net/ontologies/copyrightonto.owl#",
    "voxels": "https://voxels.com/",
    "external_link": "https://opensea.io/metadata/external_link",
    "animation_url": "https://opensea.io/metadata/animation_url",
    "youtube_url": "https://opensea.io/metadata/youtube_url"
  },
  "@id": "did:eip155:4:erc721:0x8E8B...3a97:3654...8891",
  "@type": "cro:Agree",
  "name": "Reuse license for 'Copyright Blockchain'",
  "description": "Grants the owner permission to make 'Copyright Blockchain' available at The Center neighbourhood and Vibes island from Jun 4, 2022, 4:39:31 PM to Jun 4, 2023, 4:39:31 PM",
  "external_link": "https://copyrightly.rhizomik.net/manifestations/QmfK...TRMS",
  "image": "ipfs://QmfKrv7JgGgekjtVMa...U2xXgWVaycAZWAVYQpTRMS",
  "animation_url": "ipfs://QmfKrv7JgGgekjtVMa...U2xXgWVaycAZWAVYQpTRMS",
  "cro:when": "2022-06-04T14:39:31.900Z",
  "cro:who": {
    "@id": "did:ethr:0x4:0x4cf0a8976397...0a0846540707da87212e17",
    "url": "https://copyrightly.rhizomik.net/creators/0x4cf0...2e17"
  },
  "cro:what": {
    "@type": "cro:MakeAvailable",
    "startTime": "2022-06-04T14:39:31.900Z",
    "endTime": "2023-06-04T14:39:31.900Z",
    "cro:who": {
      "owns": { "@id": "did:eip155:4:erc721:0x8E8B...3a97:3654...8891" }
    },
    "cro:what": {
      "@id": "ipfs://QmfKrv7JgGgekjtVMa...U2xXgWVaycAZWAVYQpTRMS",
      "@type": "cro:Manifestation",
      "name": "Copyright Blockchain",
      "url": "https://copyrightly.rhizomik.net/manifestations/QmfK...TRMS"
    },
    "cro:with": {
      "@id": "https://voxels.com/play"
    },
    "cro:where" : [
       { "@id": "voxels:neighborhoods/the-center" },
       { "@id": "voxels:islands/vibes" }
    ]
  }
}
\end{lstlisting}

\break

\section{Copyright Reasoning Example}

\begin{lstlisting}[language=sparql, caption={SPARQL query implementing the reasoning capabilities to respond to reuse requests based on the terms licensed by CopyrightLY's NFTs}, label=lst:sparql]
PREFIX  cro:    <https://rhizomik.net/ontologies/copyrightonto.owl#>
PREFIX  voxels: <https://voxels.com/>
PREFIX  geof:   <http://www.opengis.net/def/function/geosparql/>
PREFIX  geos:   <http://www.opengis.net/ont/geosparql#>
PREFIX  schema: <https://schema.org/>
PREFIX  vcard:  <http://www.w3.org/2006/vcard/ns#>
PREFIX  xsd:    <http://www.w3.org/2001/XMLSchema#>

SELECT DISTINCT  ?isAuthorized ?why
WHERE {
    # Inputs
    BIND(<did:ethr:0x4:0xFd9c5fC8FCecF6...fB34778E0b00C4a70f5> AS ?reuser)
    BIND(<ipfs://QmfKrv7JgGgekjtVMa...U2xXgWVaycAZWAVYQpTRMS> AS ?what)
    BIND("POINT((-65.1 1))"^^geos:wktLiteral AS ?coords)
    # Currently valid licensing NFT
    ?nft  a         cro:Agree ;
          cro:what  ?term ;
          cro:when  ?agreeDate
    FILTER ( xsd:dateTime(?agreeDate) <= now() )
    # Licensing NFT owned by reuser
    ?reuser schema:owns       ?nft .
    # Licensing NFT terms allow to make available the intended content now
    ?term   a                 cro:MakeAvailable ;
            cro:what          ?what ;
            schema:startTime  ?start
    FILTER ( now() >= xsd:dateTime(?start) )
    # Get parcel containing coordinates
    ?parcel vcard:hasGeo ?geometry
    FILTER(geof:sfWithin(?coords, ?geometry))
    # Licensed parcel or contained in licensed territory
    {   { ?term  cro:where  ?parcel }
      UNION
        { ?term  cro:where  ?territory .
          ?parcel (schema:containedIn)+ ?territory
        }
    }
    # Outputs
    BIND(bound(?nft) AS ?isAuthorized)
    BIND(?nft AS ?why)
  }
\end{lstlisting}

\end{document}